\documentclass[10pt,prl,aps,showpacs,twocolumn,superscriptaddress]{revtex4-1}

\usepackage{graphicx,bm}
\usepackage{amsmath}
\usepackage{amssymb}
\usepackage{epsfig}
\usepackage{epsf}
\usepackage{verbatim}
\usepackage[usenames]{color}
\usepackage{upgreek}
\usepackage{float}
\usepackage[breaklinks=true,colorlinks,citecolor=blue,linkcolor=blue,urlcolor=blue]{hyperref}

\newcommand{\divg}{\nabla\cdot}

\begin{document}
\title{Nonlocal spin transport as a probe of viscous magnon fluids}
\author{Camilo Ulloa}\email{c.ulloa@uu.nl}
\affiliation{Institute for Theoretical Physics, Utrecht University, Princetonplein 5, 3584 CC Utrecht,~The Netherlands}
\author{A. Tomadin}
\affiliation{Department of Physics, Lancaster University, Lancaster LA1 4YB,~United Kingdom}
\author{J. Shan}
\affiliation{Physics of Nanodevices, Zernike Institute for Advanced Materials, University of Groningen, 9747 AG Groningen,~The Netherlands}
\author{M. Polini}
\affiliation{Istituto Italiano di Tecnologia, Graphene Labs, Via Morego 30, I-16163 Genova,~Italy}
\author{B. J. van Wees}
\affiliation{Physics of Nanodevices, Zernike Institute for Advanced Materials, University of Groningen, 9747 AG Groningen,~The Netherlands}
\author{R. A. Duine}
\affiliation{Institute for Theoretical Physics, Utrecht University, Princetonplein 5, 3584 CC Utrecht,~The Netherlands}
\affiliation{Department of Applied Physics, Eindhoven University of Technology,
P.O. Box 513, 5600 MB Eindhoven, The Netherlands}

\begin{abstract}
Magnons in ferromagnets behave as a viscous fluid over a length scale, the momentum-relaxation length, below which momentum-conserving scattering processes dominate. We show theoretically that in this hydrodynamic regime viscous effects lead to a sign change in the magnon chemical potential, which can be detected as a sign change in the nonlocal resistance measured in spin transport experiments. This sign change is observable when the injector-detector distance becomes comparable to the momentum-relaxation length. Taking into account momentum- and spin-relaxation processes, we consider the quasiconservation laws for momentum and spin in a magnon fluid. The resulting equations are solved for nonlocal spin transport devices in which spin is injected and detected via metallic leads. Because of the finite viscosity we also find a backflow of magnons close to the injector lead. Our work shows that nonlocal magnon spin transport devices are an attractive platform to develop and study magnon-fluid dynamics.
\end{abstract}

\maketitle


\textit{Introduction.}---Hydrodynamics has been a universal theme across physics~\cite{LandauFluid, BatchelorFluid,ChaikinBook} due its universal applicability, and because novel systems that warrant a hydrodynamic description keep on emerging.  Recent examples of new hydrodynamic systems are strongly interacting cold-atom systems and the quark-gluon plasma~\cite{NPFsreview,QGP,cao_science_2011,elliott_prl_2014}. Very recently, it has been shown that electrons in almost defect-free solid-state conductors can reach the hydrodynamic regime where the electrons collide more frequently among each other than with phonons or impurities~\cite{Torre2015, Bandurin2016, Braem2018, Levitov2016}. In this regime, the electron viscosity becomes important and has been shown to lead, for example, to super-ballistic charge transport through point contacts~\cite{LevitovPNAS, Kumar2017}, to the possibility of measuring the Hall viscosity~\cite{Pellegrino2017,Berdyugin2018}, and, in the case of finite spin-orbit coupling, to large current-induced spin densities~\cite{Ruben}. 

The realization of viscous electron systems begs the question of whether there may be other solid-state platforms for fluid dynamics. Based on the work of Hohenberg and Halperin~\cite{Halperin}, Reiter and Schwabl answered this question affirmatively by theoretically proposing magnons, the quanta of spin waves in ferromagnets, as the entities for making up this fluid~\cite{Reiter, Michel1969, Michel1970}. 
Very recently, Prasai \textit{et al}.~\cite{PoisMagnons} have revived this direction by proposing that the observed enhancement of the magnon heat conductivity in their experiment is due to hydrodynamic Poiseuille flow of magnons. Even more recently, Rodriguez-Nieva et al.~\cite{SecondSound} have proposed to measure the second sound mode of magnons using spin qubit magnetometers~\cite{CDu2017,Toeno2015} as a probe of magnon hydrodynamics. These latter examples probe the spin transport of the magnon fluid indirectly, either through heat transport or through the existence of a hydrodynamic mode. 
\begin{figure}[H] 
	\centering
	\includegraphics[width=0.99\columnwidth]{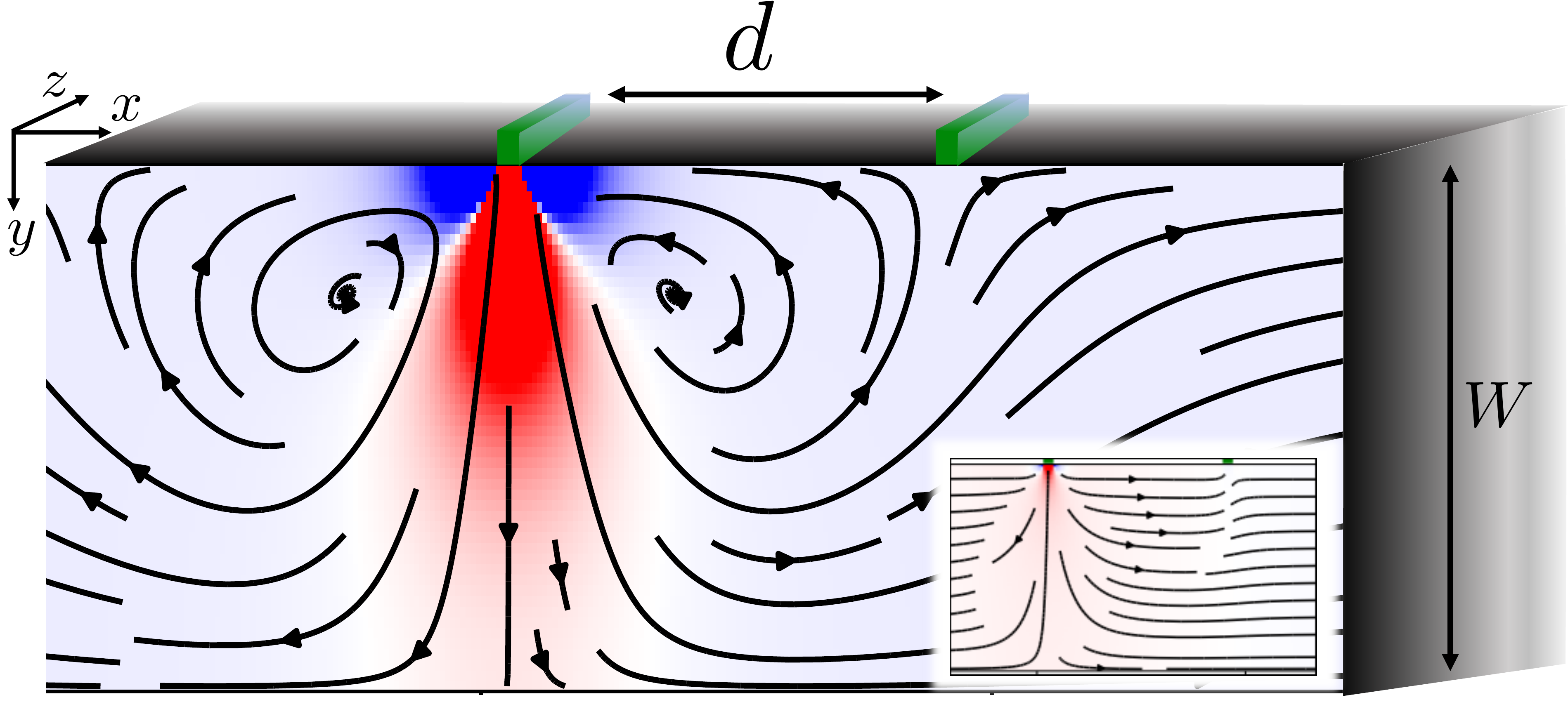}
	\caption{Schematics of a non-local transport device for the detection of viscous magnon flow. Two metallic leads of width $w_{\rm l}$ (depicted in green) are placed on top of a ferromagnetic insulator and are separated by a distance $d$. The ferromagnetic insulator has dimensions $W\times L$ in the $\hat{\bm x}$-$\hat{\bm y}$ plane while the system is translational invariant along $\hat{\bm z}$. The left lead hosts a spin accumulation ($\mu_s>0$) and injects spin into the ferromagnetic insulator. The right lead is modelled as a spin sink ($\mu_s =0$) and acts as spin detector. Due to viscous effects, the current has nonzero vorticity close to the injector, which leads to local changes in the direction of the magnon current ($\bm{j}_{\rm m}$) and to sign changes in the magnon chemical potential ($\mu_{\rm m}$). The streamlines of the magnon current are depicted with black arrows, while the color code indicates the behaviour of the normalized variations of the magnon chemical potential around its spatial average, i.e.~$[\mu_{\rm m}-\langle\mu_{\rm m} \rangle]/\mu_{\rm m}^{\rm max}$. The main panel shows the result for a viscous magnon fluid ($ D_{\nu} \sim W$) while the inset shows the result for the diffusive regime ($ D_{\nu}\to 0$). The change of sign of the spin current injected into the detector provides evidence for the existence of a viscous magnon fluid.\label{fig:setup}}
\end{figure}
In this Letter, we give theoretical evidence that non-local spin transport experiments~\cite{LudoNature} provide direct and unambiguous signatures of magnon hydrodynamics. Our proposal is motivated by recent developments in spintronics that have shown that interfacial exchange interactions enable the injection of spin from a normal metal into a magnetic insulator across an interface that separates them. The accumulation of spin in the normal metal can be addressed electrically, via the spin Hall effect~\cite{SinovaReview} and its inverse, in materials in which spin-orbit coupling is sufficiently strong. The first experiment in this direction was focused on the demonstration of the spin Seebeck effect~\cite{SpinSeebeck}, in which a thermally-driven magnon spin current across the magnetic insulator Yttrium Iron Garnett (YIG) is injected into the normal metal Pt, leading to 
an inverse spin Hall voltage~\cite{Saitoh2006}. Here, we focus on non-local devices consisting of two normal-metal leads placed on top of a magnetic insulator (see Fig.~\ref{fig:setup}), which act as reservoirs for spin injection and detection. These devices, pioneered by Cornelissen et al.~\cite{LudoNature}, have been used to probe and influence magnon spin transport across a variety of set-ups, regimes, and materials.

In this Letter, we show that the non-local voltage in spin transport experiments on devices as in Fig.~\ref{fig:setup}, i.e., the voltage measured across the detector divided by current in the injector, changes sign as the injector-detector distance becomes comparable to the momentum-relaxation length of the magnons. This relaxation length is the length scale below which magnon-fluid dynamics arising from momentum-conserving collisions manifests. Our results therefore demonstrate that measurements on non-local magnon spin transport devices are attractive to probe magnon-fluid dynamics.

\textit{Bulk magnon hydrodynamics.}---We consider a ferromagnetic insulator with its equilibrium spin pointing in the $-\hat{\bm x}$ direction.  At sufficiently low temperatures the relevant excitations are Holstein-Primakoff magnons~\cite{Holstein}, which carry $\hbar$ spin in the $\hat{\bm x}$ direction (ignoring ellipticity).  At nonzero temperature $T$ the equilibrium density $\rho_{0}$ of magnons scales like $\rho_0\propto (T/T_{\rm C})^{3/2}$, where $T_{\rm C}$ is the Curie temperature. We consider a magnon system in which momentum-conserving and magnon-conserving magnon-magnon collisions occur more frequently than momentum-non-conserving scattering processes. The latter can be both spin-conserving, e.g.~magnon-phonon interaction due to modulation of exchange, and spin-non-conserving, e.g.~magnon-phonon interaction due to modulation of anisotropy. The spin-non-conserving processes lead to decay of the magnon number. We denote by $1/\uptau_{\rm mm}$ the rate for momentum-conserving and magnon-conserving magnon-magnon interactions. The rate for momentum-non-conserving scattering processes is denoted by $1/\uptau_{\rm m}$, whereas the rate for spin-non-conserving scattering processes is denoted by $1/\uptau_{\rm mr}$. Because the latter typically contribute also to momentum relaxation, one usually has $1/\uptau_{\rm mr} \ll 1/\uptau_{\rm m}$. We will come back to estimates of the various time scales below. For the time being, we remark that since exchange dominates momentum-conserving magnon-magnon interactions and is typically the strongest interaction in ferromagnets, it is likely that the regime where $1/\uptau_{\rm mm} \gg  1/\uptau_{\rm m}$ may be reached experimentally. In this regime, effects due to magnon viscosity may manifest for sufficiently small length scales causing the magnons to behave as a fluid, as we discuss now. 

We set up a hydrodynamic theory of magnon fluids in terms of two quasi-conserved quantities, namely, momentum and spin (or magnon number). We do not consider the energy. This is primarily because we consider spin transport driven by spin accumulation in the metallic leads adjacent to the magnetic insulator. As shown in Ref.~\cite{LudoPRB2016}, such spin transport is described by introducing a magnon chemical potential rather than temperature. Moreover, discarding energy and thermal effects makes the resulting system of equations much simpler to handle. 

The quasi-conservation laws for the magnon number and momentum of the magnon fluid are given by
\begin{eqnarray}\label{eq:cons2}
\partial_{t}\rho_{\rm m} + \nabla \cdot(\rho_{\rm m} {\bm v})  =-\dfrac{\hbar}{2e}\dfrac{\sigma_{\rm m}}{\ell_{\rm m}^2}\mu_{\rm m}~,
\end{eqnarray}
and
\begin{eqnarray}\label{eq:cons1}
\rho_{\rm m}[\partial_t {\bm v} +({\bm v}\cdot\nabla){\bm v} ] = -\dfrac{\hbar}{2e}\dfrac{\sigma_{\rm m}}{\uptau_{\rm m}}\nabla {\mu_{\rm m}}\nonumber \\
+ \eta\nabla^{2}{\bm v} +\eta'\nabla(\divg{\bm v})-&\dfrac{\rho_{\rm m}}{\uptau_{\rm m}}{\bm  v}~,
\end{eqnarray}
where $\rho_{\rm m}$ is the magnon (particle) density, ${\bm v}$ the magnon velocity, $\mu_{\rm m}$ the magnon chemical potential, $\sigma_{\rm m}$ the magnon spin conductivity, and $\ell_{\rm m}$ the magnon spin diffusion length. Here $\eta'= \chi +\eta/3$, where $\eta$ is the dynamical shear viscosity and $\chi$ is the bulk viscosity that both arise from the momentum-conserving magnon-magnon interactions. The bulk viscosity $\chi$ quantifies the stress related dissipation due to time-dependent volume changes\cite{ChaikinBook}. In the stationary regime $\chi$ leads to a small renormalization of  the magnon spin diffusion length and the magnon spin conductivity which we neglect from now on~\cite{BulkVisc}. Eq.~\eqref{eq:cons2} is the magnon continuity equation, which is augmented to include magnon decay. Eq.~\eqref{eq:cons1} is the Navier-Stokes equation including a phenomenological term $-\rho_{\rm m} {\bm v}/\uptau_{\rm m}$ due to relaxation of momentum. It also includes a term corresponding to the effective force $\propto \nabla {\mu_{\rm m}}$ on magnons (note that $\sigma_{\rm m} \propto \uptau_{\rm m}$). 

Throughout this Letter, we consider the above equations in the linearized and time-independent (i.e.~steady-state) regime so that the magnon fluid is described by
\begin{eqnarray}
\label{eq:contlin}
\nabla\cdot {\bm j}_{\rm m} =-\frac{\hbar \sigma_{\rm m}}{2e\ell_{\rm m}^2}\mu_{\rm m}\label{eq:Nav1}
\end{eqnarray}
\begin{eqnarray}
D_{\nu}^2\nabla^{2}{\bm j}_{\rm m}-\dfrac{\hbar\sigma_\nu}{2e}\nabla\mu_{\rm m} = {\bm j}_{\rm m}\label{eq:Nav2}~,
\end{eqnarray}
where ${\bm j}_{\rm m} =\rho_0 {\bm v}$ is the linearized magnon current with $rho_0$	 the average magnon density,  $D_\nu=\sqrt{\nu \uptau_{\rm m}}$ is the momentum-relaxation length or, equivalently in the present isotropic case, the vorticity diffusion length~\cite{Vort_diffusion}, $\nu= \eta/\rho_0$ is the kinematic viscosity, and $\sigma_\nu = \sigma_{\rm m}(1+D_{\nu}^2/3\ell_{\rm m}^2)$ the magnon spin conductivity that includes viscous effects. A straightforward calculation using Eqs.~\eqref{eq:Nav1}-\eqref{eq:Nav2} yields a diffusion equation for the chemical potential $\nabla^2 \mu_{\rm m} = \mu_{\rm m}/\ell_\nu^2$, where $\ell_\nu = \ell_{\rm m}(1+4D^2_{\nu}/3\ell_{\rm m}^2)$ is the magnon spin diffusion length that includes viscous effects. In the absence of viscosity, $\nu=0$, we have that $\ell_{\nu}=\ell_{\rm m}$, $\sigma_{\nu}=\sigma_{\rm m}$, and ${\bm j}_{\rm m} = -\hbar\sigma_{\rm m}/ 2e  \nabla\mu_{\rm m}$ as expected. The presence of viscosity together with momentum relaxation leads to the length scale $D_\nu$. In the absence of viscosity we recover the diffusive regime~\cite{LudoPRB2016}. The magnon number decay leads, in addition to a finite $\ell_{\rm m}$, also to a finite compressibility of the magnon fluid, since $\nabla\cdot {\bm j}_{\rm m} \neq 0$.

\textit{Non-local magnon transport.}---We consider a non-local device as depicted in Fig.~\ref{fig:setup}, consisting of two metallic leads on top of a ferromagnetic insulator. The left lead hosts a nonzero spin accumulation $\mu_s\left. \right|_{\rm inj} = \mu_\uparrow-\mu_\downarrow$ that is pointing in the $\hat{\bm x}$-direction. Experimentally, this is typically generated by the spin Hall effect. The right lead is the detector and is treated as an ideal spin sink, i.e.~$\mu_s\left. \right|_{\rm det} = 0$.
\begin{figure}[t] 
\centering
\includegraphics[width=0.99\columnwidth]{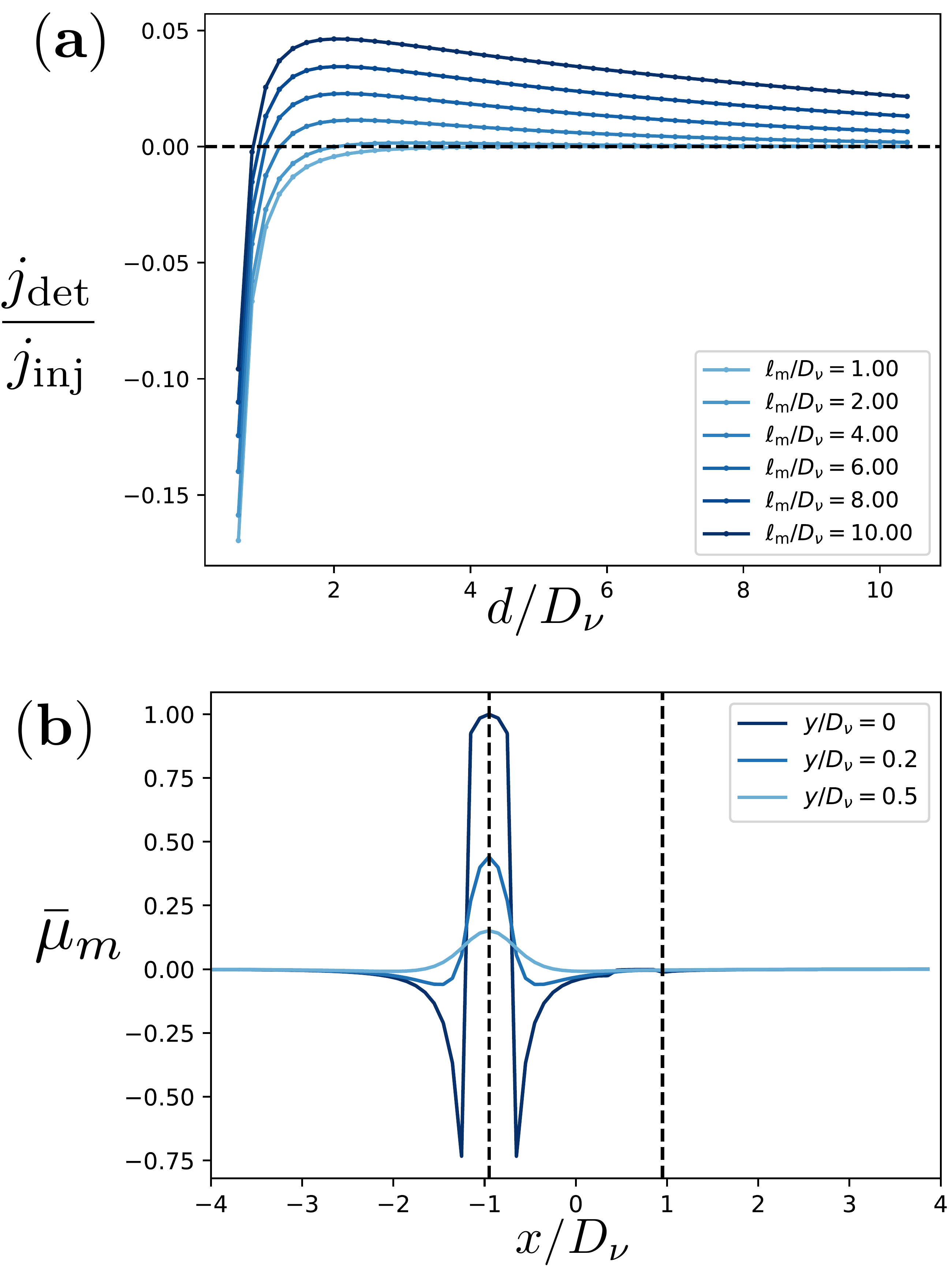}
\caption{\textbf{(a)}~Non-local signal as a function of injector-detector distance $d$ for a sample of width $W=5~D_\nu$ considering different values of the magnon spin diffusion length.  \textbf{(b)}~Profile of the normalized magnon chemical potential $\bar{\mu}_{\rm m}= \mu_{\rm m} /\mu_{\rm m}^{\rm max} $  along the $\hat{\bm x}$-direction at different distances from the upper boundary [see Fig.~\ref{fig:setup}]. We consider $d=1.6~D_\nu$ and $\ell_{\rm m} / D_{\nu}=1$. The dashed lines depict the central position of the injector (left) and detector (right). \label{fig:results}}
\end{figure}
The injected magnon current from (or into) the interface with the lead, in the linear response regime, depends on the difference between the spin accumulation and the magnon chemical potential, i.e.~$\left. {\bm j}_{\rm m}\cdot \hat{\bm n}\right|_{\rm int} = g_{\rm s} (\mu_s - \left.\mu_{\rm m})\right|_{\rm int}$, where $g_{\rm s}$ is the interfacial spin conductance~\cite{LudoPRB2016}, and $\hat{\bm n}$ is a unit vector normal to the lead-ferromagnet interface. At all boundaries where there is neither an injector nor a detector---which we term open boundaries (OB)---the normal component of the current vanishes ($\left.{\bm j}_{\rm m}\cdot \hat{\bm n}\right|_{\rm OB}=0$). The boundary conditions for the tangential component of the current are characterized by a phenomenological slip-length $\ell_{\rm b}$~\cite{Torre2015,Pellegrino2017}. This quantifies the loss of momentum of magnons moving parallel to the boundary. If $\ell_{\rm b} \to 0$ the component of the magnon current parallel to the boundary vanishes at the boundary leading to a no-slip condition. In the opposite regime ($\ell_{\rm b} \to \infty$), magnons can be thought of as slipping along the boundary with zero friction i.e. without exerting stress on the boundary. The equation that encodes this behaviour is 
$\ell_{\rm b}(\partial_y j_x + \partial_x j_y\left. ) \right |_{\rm OB} = \left.{\bm j}_{\rm m} 
\cdot \hat{\bm t}\right |_{\rm OB}$, where $\hat{\bm t}$ is a unit vector tangent to the OBs, such that $\hat{\bm n} \times \hat{\bm t} = \hat{\bm z}$. 

\textit{Results.}---We assume translational invariance along the $\hat{\bm z}$ direction and solve Eqs.~\eqref{eq:Nav1}-\eqref{eq:Nav2} in the $\hat{\bm x}$-$\hat{\bm y}$ plane, applying the boundary conditions mentioned above. Due to the coupling of the magnon current and the magnon chemical potential in Eqs.~\eqref{eq:Nav1}-\eqref{eq:Nav2}, as well as in the boundary conditions, finding an analytical solution of our system of partial differential equations is, to the best of our knowledge, not readily possible. We therefore resort to numerics. The solution for the current streamlines is depicted by black arrows in Fig.~\ref{fig:setup}, both in the presence and in the absence of viscosity (main panel and inset, respectively). The color code in Fig.~\ref{fig:setup} shows the behaviour of the normalized variations of the magnon chemical potential around its spatial average, i.e.~$[\mu_{\rm m}-\langle\mu_{\rm m} \rangle]/\mu_{\rm m}^{\rm max}$. We focus our analysis on the ratio between detected and injected magnon currents
\begin{equation}
\dfrac{j_{\rm det}}{j_{\rm inj}}=\dfrac{\int_{S_{\rm det}} (\left.{\bm j}_{\rm m}\cdot \hat{\bm n})\right|_{\rm det}}{\int_{S_{\rm inj}} (\left.{\bm j}_{\rm m}\cdot \hat{\bm n})\right|_{\rm inj}}~,\label{eq:ratio}
\end{equation}
where the integration is over the interfaces between the ferromagnetic insulator and the metallic leads. 
Assuming both leads have the same spin Hall angle and spatial dimensions, this ratio is proportional to $R_{\rm nl}/R_{0}$, where $R_{\rm nl}$ is the non-local resistance that is measured experimentally, i.e.~the voltage across the detector divided by the current through the injector, and where $R_0$ is the resistance of the leads. 
In non-local transport experiments involving YIG as the magnetic insulator and Pt strips as leads, the values of the parameters and dimensions of the device are of the following order of magnitude~ \cite{LudoPRB2016}: $D_{\nu}\sim 0.1~\mu\text{m}$, $\sigma_{\rm m} \sim 10^{5}~\text{S/m}$, $g_s\sim 10^{13}~\text{S/m}^2$, $w_{\rm l}\sim 0.5~D_{\nu}$, and $\mu_s\sim 9~\mu\text{V}$. We consider these values in our numerical results. We fix the length of the device to $L = 50~D_{\nu}$.
We analyze the limit $\ell_{\rm b}\to\infty$ (our results do not change qualitatively if $\ell_{\rm b}\to 0$).
We observe a change in the sign of $\mu_{\rm m}$ as the injector-detector distance is decreased, which occurs only in the presence of viscosity and corresponds to a depletion of magnons. The result of the ratio defined in Eq.~\eqref{eq:ratio} is shown in Fig.~\ref{fig:results}~{\bf (a)} for different values of $\ell_{\rm m}$. The change of sign in the chemical potential is another remarkable consequence of the viscous effects and is the ultimate responsible of the negative signal. Different profiles of the chemical potential are shown in Fig.~\ref{fig:results}~{\bf (b)}.
The change of sign disappears in absence of viscosity and we recover the results of Ref.~\onlinecite{LudoNature} in the diffusive regime. 

The negative sign can be understood as a magnon current flowing from the detector into the sample. The current is dragged out of the detector because of the nonzero viscosity, in rough analogy with the profile of the flow in the Venturi effect.

The sign change discussed above is the main result of this Letter, and gives a direct and unambigous signature of the magnon fluid regime, which can be readily probed in currently available devices. We remark that in the limit $\ell_{\rm m}/D_{\nu}\to \infty$ our results are analogous to those obtained for hydrodynamics of electrons in graphene~\cite{Bandurin2016}. 

We also observe the presence of whirlpools in the magnon flow accompanying the sign change. We considered two limiting cases for the thickness of the magnetic insulator. Namely, $W\ll \lbrace D_\nu,\ \ell_{\rm m}\rbrace$, and $W\gg \lbrace D_\nu,\ \ell_{\rm m}\rbrace$, which we call the thin-film and thick-film regime, respectively. In the thin-film limit the whirlpools only appear when $D_\nu\sim\ell_{\rm m}$. In the thick-film regime, the whirlpools are only present when $D_{\nu}>\ell_{\rm m} $.

\textit{Discussion.}---We have shown that non-local spin transport measurements are able to probe the existence of the viscous regime of magnon hydrodynamics. Let us now estimate the parameters for YIG. Following Ref.~\cite{LudoPRB2016}, we have that at room temperature $\uptau_{\rm m} \sim \uptau_{\rm mp}\sim 1$-$100~{\rm ps}$, $\uptau_{\rm mr} \sim 1~{\rm ns}$, and $\uptau_{\rm mm} \sim 0.1$-$1~{\rm ps}$. These estimates show explicitly how these numbers fulfill the required conditions for the hydrodynamic regime, and how YIG at room temperature may well be in the regime where viscous effects are important. Moreover, since $1/\uptau_{\rm mm}$ and $1/\uptau_{\rm mp}$ have very different dependence on temperature, it is likely that the hydrodynamic regime may be reached for sufficiently clean YIG as a function of temperature. Note that these estimates are supported by the results of Ref.~\onlinecite{SecondSound}.

The above-mentioned time scales, however, are averages over all magnon modes dominated by modes around the thermal energy, and we have tacitly assumed that the momentum-conserving magnon-magnon interaction rate is the fastest process for all magnon modes. At elevated temperature, this assumption is likely to be correct as for magnons with energy around the thermal energy the magnon-magnon interactions are dominated by the ex- change interactions which are typically strong.

We approximate the kinematic viscosity 
$\nu= v_{\rm th}^2\uptau_{\rm mm}/3\sim (0.5$-$5)\times 10^{-5}~{\rm m}^2/{\rm s}$, which leads to $D_{\nu}\sim  1$-$100~{\rm nm}$. 
While being small, this length scale is not beyond reach of experiments. Moreover, since both $\nu$ and $\uptau_{\rm mp}$ are expected to strongly increase upon lowering the temperature, $D_{\nu}$ is expected to increase upon lowering the temperature as well. 

We have solved our hydrodynamic equations in the linear response regime. We now assess if the non-linear turbulent regime may be reached with non-local devices based on YIG and Pt. With a simple rescaling of Eqs.~\eqref{eq:cons2}-\eqref{eq:cons1} we define the Reynolds number, which governs the relative importance of non-linear effects, as ${\rm Re} =  e\mu_{\rm m} W\uptau_{\rm m}/  (k_{\rm B} T\ell_{\rm m}\uptau_{\rm mm})$ which, considering $\mu_{\rm m}\sim 1~\mu {\rm V}$, seems to be larger than unity at room temperature for $W>\ell_{\rm m}$. To assess the importance of non-linear effects more quantitatively, future work should focus on stability analysis of the linear solutions that we obtained.  Another direction of research would be to extend our model to include energy conservation, which encodes thermal transport effects.

It would be straightforward to include energy transport on top of the spin transport. In that case Eq.\eqref{eq:Nav2} needs to be complemented with a term $\sim \nabla T$ and a continuity-like equation [similar to Eq.~\eqref{eq:Nav1}] for the heat current needs to be considered. In a real experiment the thermal and spin contributions can be discriminated by analysing the harmonics of the measured signal using lock-in techniques. The contribution arising from the electrically injected magnons corresponds to the first harmonic of the measured signal, while the thermally injected magnons are related with its second harmonic \cite{LudoNature}.  

The slip length $\ell_{\rm b}$ is another interesting quantity that can be explored in more detail. In particular, in this work the slip length does not play a relevant role but in devices with narrow ``bottlenecks" of ferromagnetic insulator it could be possible to obtain strong Poiseuille flow of magnons in the no-slip regime ($\ell_{\rm b}\to 0$). In principle, the slip length can be tuned by means of composition of adjacent materials and interface, leading to changes in momentum transfer processes across the interface. 

The experimental verification of our proposal in YIG is challenging to realize mainly due to the length scales below where viscous effects are dominant ($D_{\nu}\sim 1-100~\text{nm}$), because placing two Pt leads on top of YIG separated by distances below $D_{\nu}$ is not easy to achieve.

The results in our Letter hopefully show that non-local devices are an attractive platform for studying magnon-fluid dynamics and for exploring, among others, the above-mentioned directions for future experimental and theoretical research. 
  
\textit{Acknowledgements.}---It is a pleasure to thank Gerrit Bauer and Iacopo Torre for
enlightening discussions. R. A. D. is a member of the DITP consortium, a program of the Netherlands Organisation for Scientific Research (NWO) that is funded by the Dutch Ministry of Education, Culture and Science (OCW). This work is in part funded by the Stichting voor Fundamenteel Onderzoek der Materie (FOM). R. A. D. also acknowledges the support of the European Research Council. Professor B. J. van Wees acknowledges the financial support from the Zernike Institute for Advanced Materials (ZIAM), and from the Nederlandse Organisatie voor Wetenschappelijk Onderzoek (NWO) through the Spinoza prize.

\bibliographystyle{elsarticle-num}

\begin{thebibliography}{100}
\bibitem{BulkVisc}The presence of bulk viscosity ($\chi$) modifies the enhanced magnon spin	 conductivity $\sigma_{\nu} \to \sigma_{\rm m} [1 + \eta'\uptau_{\rm m}/\rho_0\ell_{\rm m}^{2}] $ and the enhanced magnon spin diffusion length $\ell_{\nu} \to \ell_{\rm m }^{2}[1 +D_{\nu}^2/\ell_{\rm m}^{2} +\eta' \uptau_{\rm m}/\rho_0\ell_{\rm m } ^2   ] $. If $\chi=0$ we recover the results shown in the main text.
\bibitem{Vort_diffusion} It is straightforward from Eqs.\eqref{eq:Nav1}-\eqref{eq:Nav2} to obtain $D_\nu^2\nabla^2\boldsymbol{\omega} = \boldsymbol{\omega}$, where $\boldsymbol{\omega}=\nabla\times{\bm j}_{\rm m}$ is the vorticity. Therefore $D_\nu$ can be interpreted as the characteristic diffusion lenght of vorticity.
\end{thebibliography}

\end{document}